\documentclass[conference]{IEEEtran}
\IEEEoverridecommandlockouts
\usepackage{cite}
\usepackage{amsmath,amssymb,amsfonts}
\usepackage{algorithmic}
\usepackage{graphicx}
\usepackage{textcomp}
\usepackage{xcolor}
\usepackage{url}
\usepackage{acronym}
\usepackage{hyperref}
\usepackage{subcaption}
\usepackage{float}

\acrodef{ABM}{Agent-Based Model}
\acrodef{MS}{Modeling and Simulation}
\acrodef{PADS}{Parallel and Distributed Simulation}
\acrodef{LoD}{Level of Detail}
\acrodef{ODE}{Ordinary Differential Equation}
\acrodef{SEIR}{Susceptible, Exposed, Infected, Recovered}
\acrodef{LUNES}{Large Unstructured NEtwork Simulator}

\makeatletter
\newcommand{\linebreakand}{%
  \end{@IEEEauthorhalign}
  \hfill\mbox{}\par
  \mbox{}\hfill\begin{@IEEEauthorhalign}
}
\makeatother

\makeatletter
\def\ps@IEEEtitlepagestyle{%
\def\@oddfoot{\mycopyrightnotice}%
\def\@evenfoot{}%
}
\def\mycopyrightnotice{%
}

\begin{document}

\title{Multilevel Modeling as a Methodology for the Simulation of Human Mobility}

\author{\IEEEauthorblockN{Luca Serena}
\IEEEauthorblockA{\textit{Department of Computer Science and Engineering}\\
\textit{University of Bologna}\\
Bologna, Italy\\
luca.serena2@unibo.it}
\and
\IEEEauthorblockN{Moreno Marzolla}
\IEEEauthorblockA{\textit{Department of Computer Science and Engineering} \\
\textit{University of Bologna}\\
Cesena, Italy\\
moreno.marzolla@unibo.it}
\linebreakand
\IEEEauthorblockN{Gabriele D'Angelo}
\IEEEauthorblockA{\textit{Department of Computer Science and Engineering} \\
\textit{University of Bologna}\\
Cesena, Italy\\
g.dangelo@unibo.it}
\and
\IEEEauthorblockN{Stefano Ferretti}
\IEEEauthorblockA{\textit{Department of Pure and Applied Sciences}\\
\textit{University of Urbino ``Carlo Bo''}\\
Urbino, Italy\\
stefano.ferretti@uniurb.it}%
\thanks{\color{red}\bf This is the author’s version of the article:
  ``Luca Serena, Moreno Marzolla, Gabriele D'Angelo, Stefano Ferretti,
  Multilevel Modeling as a Methodology for the Simulation of Human
  Mobility, proc. 2022 IEEE/ACM 26th International Symposium on
  Distributed Simulation and Real-Time Applications (DS-RT'22), Alès,
  France, September 26--28, 2022, pp. 49--56''.  \textcopyright 2022
  IEEE. Personal use of this material is permitted.  Permission from
  IEEE must be obtained for all other uses, in any current or future
  media, including reprinting/republishing this material for
  advertising or promotional purposes, creating new collective works,
  for resale or redistribution to servers or lists, or reuse of any
  copyrighted component of this work in other works. The publisher
  version of this paper is available at
  \url{https://doi.org/10.1109/DS-RT55542.2022.9932080}.}  }

\maketitle

\begin{abstract}
Multilevel modeling is increasingly relevant in the context of
modelling and simulation since it leads to several potential benefits,
such as software reuse and integration, the split of semantically
separated levels into sub-models, the possibility to employ different
levels of detail, and the potential for parallel execution. The
coupling that inevitably exists between the sub-models, however,
implies the need for maintaining consistency between the various
components, more so when different simulation paradigms are employed
(e.g.,~sequential vs parallel, discrete vs continuous). In this paper
we argue that multilevel modelling is well suited for the simulation
of human mobility, since it naturally leads to the decomposition of
the model into two layers, the "micro" and "macro" layer, where
individual entities (micro) and long-range interactions (macro) are
described. In this paper we investigate the challenges of multilevel
modeling, and describe some preliminary results using prototype
implementations of multilayer simulators in the context of epidemic
diffusion and vehicle pollution.
\end{abstract}


\section{Introduction}\label{sec:introduction}

Multilevel modeling is a methodology that refers to the hierarchical
decomposition of a system into multiple, cooperating models.  This
approach received increasing interest in recent years, due to the need
to create scalable modeling and simulation solutions devoted to the
study of complex systems, that in many situations require a high level
of detail and are composed of a large number of
entities~\cite{gda-simpat-iot}.  Multilevel modeling is often referred
to as \emph{multilayer modeling} or simply \emph{hierarchical
  modeling}. Due to the lack of formal definitions, through this paper
we will use the terms \emph{multilevel} or \emph{multilayer} modelling
interchangeably, to denote hierarchical models\footnote{We also allow
  models with one level, i.e., "flat" hierarchies.} where:

\begin{itemize}
\item sub-models can be of different types (e.g.,~continuous, discrete
  and/or hybrid models);
\item sub-models may have a different level of detail, e.g., in terms
  of spatial or temporal resolution; sub-models are allowed to change
  the level of detail at run time.
\end{itemize}

The advantages and limitations of hierarchical modeling and simulation
have already been studied in the past~\cite{718100}, mainly in the
context of sequential models (more details will be provided
shortly). Hierarchical modeling is based on one of the cornerstones of
computer science, that is, the principle of decomposition. Breaking
complex entities into smaller pieces makes the system easier to build
and understand.

Decomposition of a complex model also brings another, less obvious
advantage: the different sub-models can be executed by independent
simulators. Different paradigms (e.g.,~continuous, event-driven
discrete, time-stepped discrete, and so on) can coexist both on
different layers and/or on different models of the same layer.
Additionally, some of the models might be executable in parallel,
allowing the application of~\ac{PADS} techniques~\cite{new-trends}.

Multilevel~\ac{MS} involves more than just decomposition: in fact, the
possibility of employing different levels of detail plays an even more
important role. The concept of~\ac{LoD} is model-specific, but
typically refers to spatial or temporal resolution, and/or the amount
of state variables that are used to encode the state of a
(sub-)model. \emph{Spatial} and \emph{temporal resolution} are the
density of (simulated) space and time subdivisions, respectively; the
idea of tuning the spatial or temporal resolution is widely used in
physics and engineering to study continuous systems such as weather
patterns, air flow around vehicles, or the diffusion of heat inside a
combustion engine, just to name a
few~\cite{multiscale-modeling}. Continuous phenomena must be
discretized in order to be solved numerically, and a finer subdivision
in space and/or time usually (but not necessarily) leads to more
accurate results.  The amount of \emph{state variables} is another
factor impacting the accuracy of a model. For example, when studying
the diffusion of an epidemic we might take into account either the
number of susceptible, infected and recovered individuals, or we might
model each person as an autonomous agent with a complex behavior that
depends on age, occupation, residence, and so forth. In the former
case, we get a coarse but very compact representation of the system
using just a handful of scalar values; in the latter case, we get a
more detailed model requiring a considerably larger state space.

It is obvious that always using the maximum~\ac{LoD} can be
computationally impractical. It is also evident that choosing the
"right" \ac{LoD} can be difficult or even impossible. For example,
models that involve human interactions, such as those studying the
diffusion of a contagious disease, tend to be highly sensitive to the
population density within an
area~\cite{tarwater2001effects}. Therefore, it seems appropriate to
use a finer spatial subdivision in densely populated areas, and a
coarser subdivision in sparsely populated ones. Unfortunately, in some
scenarios the population density may change over time, so that it is
impossible to predict exactly where a finer subdivision is
required. In these situations, the possibility of dynamically changing
the~LoD as the simulation evolves is highly desirable.

Multiscale models naturally allow the use of different~\acp{LoD} at
the different levels; sub-models can also dynamically tune
their~\ac{LoD} to effectively focus on interesting local phenomena,
with no or minimal impact on the rest of the simulator.

Despite the advantages discussed above, decomposition brings with it
some issues that need to be addressed. First of all, one needs to
decide how the system should be split into modules, and how the
modules interact. Interactions are particularly problematic as they
may impose a significant overhead: if a system is split into~$N$
modules, and each module needs to talk to each other, the number of
interactions is~$O(N^2)$ so that doubling the number of modules will
increase the number of interactions four-fold.

Also, the use of a multilevel methodology requires extra effort for
achieving consistency among models. A first problem arises when
continuous and discrete simulators need to interact. One of the
problems is the conversion between continuous and discrete variables,
which can lead to known issues. An example is when discrete variables
describing a population are provided as input for a continuous model
that manipulates such values and then retrieves again a discrete
output. During the cast operations, a loss could occur due to the
discretization of the variables, leading to a population loss that can
be more or less significant depending on the size of the population.

In this paper we assess whether multilevel models are suitable for
analyzing scenarios involving human mobility. This application area
has been selected due to its increasing relevance for studying such
diverse scenarios as the propagation of contagious diseases,
quantifying the environmental impact of different policies encouraging
smart mobility, better planning of transportation systems to improve
efficiency, reduce costs and reduce pollution, and so on. To this aim,
in Section~\ref{sec:background} we review some related scientific
literature. In Section~\ref{sec:multilevel} we describe in detail a
multilevel modeling paradigm that will be applied to two simple case
studies: the study of epidemics (Section~\ref{sec:epidemic}) and smart
transportation systems (Section~\ref{sec:mobility}). The case studies
are proof-of-concepts, intended to test the feasibility of the
proposed approach. Finally, conclusions and future research directions
will be illustrated in Section~\ref{sec:conclusions}.

\section{Background and Related Works}\label{sec:background}

Multilevel modeling has been largely used in the field of modelling
and simulation. The SHARPE tool~\cite{sharpe} is an early -- and still
widely used -- software package that enables multiple modeling
notations to be used to describe a complex system. SHARPE mainly deals
with formal notations such as block diagrams, reliability graphs,
queueing networks, Markov chains, Petri nets and so forth. It allows
the users to choose the number of levels, the type of model at each
level, and how information is exchanged between levels, to perform
analyses about the reliability and availability of large
systems~\cite{sahner1987reliability}.

The areas of application for multilevel modelling are very wide,
ranging from chemical and biological investigations to scenarios
linked with mobility and human activities.  In recent years, due to
the unfortunate COVID-19 outbreak, a lot of studies were carried out
in the epidemiological field.  In~\cite{martcheva2015coupling}
within-host and between-host models are coupled, with the former
describing the evolution of the disease at the cell level and the
latter depicting contagion dynamics.  Other works employ one level to
model human mobility, while another one is in charge of describing the
evolution of the outbreak in the local areas. These models enable to
simulate the diffusion of pathogens outside of the source place. For
example, in~\cite{wang2013impact} a model represents the bidirectional
recurrent commuting ﬂows that couple two populations, while SIR
(i.e.,~compartmental model where people are either susceptible,
infected or recovered) describes the local evolution of the
epidemic. In the proposed approach, each individual was characterized
by a contact rate typical of the belonging subpopulation, thus
enabling to take into account the heterogeneity of the characteristic
contact rates in diﬀerent subpopulations.  Similarly,
in~\cite{chang2021mobility} dynamics of COVID-19 spread are
investigated by using a bipartite graph, composed of census block
groups (CBG, i.e.,~geographical units typically containing some
thousands of individuals) and points of interests (POI), where the
weights of the edges indicate the number of visitors from a CBG to a
POI based on real anonymized data.  Again regarding epidemiological
scenarios, in~\cite{peng2021multilayer} one layer is employed to model
the spread of opinion regarding social distancing rules, with
individuals either being (i)~in favour of restrictions, (ii)~averse to
social distancing limitation or (iii)~uninformed. Uninformed
individuals according to the model are potentially influenced by both
factions, and choose their opinion depending on their social
connectivity. This level is directly coupled with the epidemic level,
because the number of people actually respecting restrictions strongly
affects the infection rate.

Another area where a multiscale approach is frequently used is human
mobility, with typical investigations being linked with urban traffic,
pedestrian mobility or people's access to public events.  In these
scenarios, micro models describe the movements of single individuals,
while macro models consider the distribution of people or vehicles
over a larger space~\cite{cristiani2011multiscale}. A common
methodology is to describe the macro and the micro behaviour with
partial differential equations and~\acp{ODE}, respectively. However,
agent-based models might be used to depict the behaviour of single
individuals, characterizing them with specific features.  For
instance, in a traffic simulation, a micro model would be in charge of
representing how a car reacts to preceding vehicles, with actions like
accelerating, braking or steering influencing the behaviour of the
other actors~\cite{olstam2004comparison}. A macro model, on the other
hand, would deal with the traffic flow, employing measures such as
density (number of vehicles per unit road length at any instant of
time), space mean speed (the average speed of the vehicles in a
certain road section), and flow (number of vehicles passing through a
point in a certain amount of time). Such frameworks could either be
used to study solutions for improving traffic circulation, or for
investigating smart city and smart transportation services such as car
sharing applications.  In~\cite{ni2011multiscale} the microscale is
represented by a car-following model flanked by a model representing
driver’s lateral control, while the macroscale traffic flow is defined
by a system of partial differential equations. In between, there is an
additional scale (i.e.,~mesoscale) describing with a distribution
function the probability of having a vehicle within certain space
ranges and speed ranges at a given time.  Another example
is~\cite{ekyalimpa2016combined}, where the intersections between
trains tracks and pedestrian spaces are simulated following a
discrete-continuous simulation approach. In this work, continuous
models were employed to mimic the train travel, its transition between
zones and the related scanning operations, while a discrete approach
was used to simulate the movements of pedestrians. 
considers behavioural rules of both trains and pedestrians, simulating
the strategy applied by the train to avoid incidents and the behaviour
of the people in response of the arrival of the train.

\section{Multilevel Modeling}\label{sec:multilevel}

Multilevel modelling is a methodology that allows complex models to be
built hierarchically. Each node of the hierarchy has some form of
control over the nodes at lower level, that may range from simple
orchestration to more complex scenarios where each level has a
different "view" of the system based on a different level of detail
(some actual examples will be given in
Section~\ref{sec:case-studies}).

\begin{figure}[ht]
\centering\includegraphics[width=\columnwidth]{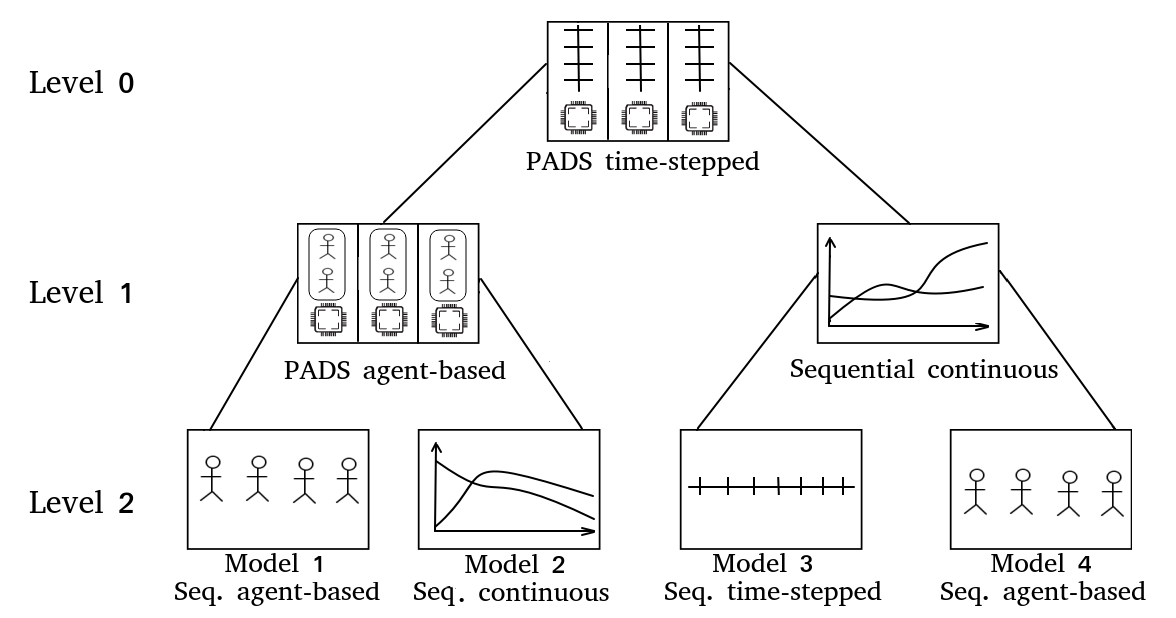}
\caption{Multilayer model using different types of sub-models and execution policies.}\label{fig:multilayer}
\end{figure}

Figure~\ref{fig:multilayer} shows a schematic representation of a
multilevel model. Different types of models can coexist in the
hierarchy (e.g., time-stepped, agent-based, continuous), as well as
different execution policies (sequential or parallel/distributed). The
role and type of each component of the hierarchy, and the structure of
the whole hierarchy, are model-specific. In general, we expect that
sub-models that are lower in the hierarchy (i.e., towards the leaves)
are more detailed than those at the top (towards the root). It may
also be possible that intermediate nodes act as pure coordinators of
their children, for example executing them in parallel if
possible. The hierarchy does not need to be static, either: a node
might spawn sub-models when needed, for example to "focus" on some
interesting emerging pattern that requires detailed study.

We can therefore classify each sub-model across the following three dimensions:
\begin{itemize}
\item \emph{Type of model} (continuous, discrete, mixed): this
  attribute refers to the representation of time.
\item \emph{Execution policy} (sequential or parallel): this attribute
  tells how a sub-model is executed.
\item \emph{Level of detail} (low, high, adaptive): this attribute
  refers to the amount of state variables that are used to represent
  the state of a sub-model.
\end{itemize}

\paragraph{Advantages}
Multilevel models provide various benefits. A complex model built upon
independent and reusable blocks might be faster to develop. From the
software engineering point of view, each building block can be
developed independently, or taken from a library of existing models
that have already been validated, therefore saving considerable time
(however, integration tests still need to be performed).

A multilevel structuring favors the parallel execution of sub-models
that have no inter-dependencies.  Simulation of spatially-located
entities (\emph{situated agents}) naturally leads to some form of
spatial decomposition, where the simulated space is partitioned across
different simulators. If the partitions are independent, some "cheap"
parallelism can be achieved by simply running the sub-models on
different execution units (processors or cores). Problems arise if
entities, e.g., those situated along the border of different
partitions, need to interact. In these cases, it is sometimes possible
to let the higher-level model take care of these
interactions. Individual sub-models might also be natively capable of
parallel execution, e.g., because they have been built upon a
parallel/distributed framework~\cite{fujimoto}.

\paragraph{Challenges}
Multilevel modeling poses several challenges, some of which are still
being investigated by the M\&S research community. First and foremost,
identifying an "optimal" partitioning -- for some suitable definition
of "optimal" -- can be difficult. As already said, models that exhibit
a spatial structure naturally lead to partitioning the simulated space
into connected, non-overlapping regions handled by different
sub-models. Even in those simple scenarios, the partitioning problem
remains nontrivial~\cite{gda-simpat-2017}.

The interaction among different models is a major issue. Although
standard interfaces for interoperation across simulators have been
proposed~\cite{HLA}, they are quite large and cumbersome to
implement. A more lightweight approach is to use wrappers file to
(i)~schedule the execution of the various components, (ii)~manage the
I/O operations and the exchange of information among the various tools
and (iii)~ensure consistency of data and state variables.

During the execution of a model, it is often necessary to use one or more streams of (pseudo-)random values. It is common practice to generate pseudo-random numbers from one initial seed to ensure the repeatability of the results. In a parallel or distributed setting, each model has its own random stream. It is therefore necessary to ensure that (i)~the random streams do not overlap, i.e., the initial seeds (or the pseudo-random generator) are chosen so that the random streams are independent, and (ii)~the result of the multilevel model is not affected by the order in which the sub-models are executed.

\paragraph{Implementation}
We now describe a possible realization of multilevel~\ac{MS}, focusing
our attention on human mobility. The choice of this application area
is motivated by its increasing relevance, e.g., to study the
propagation of infectious diseases, to investigate better and
"smarter" transportation systems, or to fight pollution by
reorganizing the way people commute to work or study.

Traditionally, human mobility has been studied using either
agent-based or continuous models based
on~\acp{ODE}~\cite{barbosa18}. Agent-based models allow developers to
accurately describe the behaviour of the actors
involved. \ac{ODE}-based models describe the aggregate behavior of a
possibly large number of agents by taking into consideration "average"
behaviors. \ac{ODE}-based models are simpler (and possibly less
accurate, particularly when there is a non-negligible probability that
agents deviate from the average behavior), and can be evaluated much
more efficiently because their performance is independent from the
number of individuals.

The easiest way to implement an agent-based simulation is to manage
the time discretely, in order to facilitate the coordination and the
interactions among the agents, which occur in certain time-steps.  On
the other hand, \ac{ODE} models are continuous. Coupling discrete and
continuous models can be problematic; we employ the standard solution
of considering the execution of the continuous sub-models as time
intervals that stretch inside the sorted list of events of the
discrete models. If a discrete model~$D$ has a continuous model~$C$ as
a sub-model, then~$D$ starts the execution of~$C$ when needed (say, at
simulated time~$T_n$). $C$ is configured to compute its state up to a
simulated time that does not exceed the instant~$T_{n+1}$ of the next
event scheduled for~$D$. Therefore, the result of the sub-model~$C$ is
made available to~$D$ at its next time step. A similar mechanism is
used if~$C$ has~$D$ as sub-model: the caller executes the callee up to
a future time that does not exceed the minimum of the duration of the
continuous phenomena and the time instant of the next discrete event.

To demonstrate the feasibility of a multilevel approach in the context
of human mobility, we have developed a prototype framework based
on~\acs{LUNES}~\cite{DANGELO202230}, NetLogo~\cite{tisue2004NetLogo}
and custom-built compartmental models based on~\acp{ODE}. In our
prototype implementations, we use~JSON-formatted data stored in
temporary files to exchange information between models.

The~\acf{LUNES} is a time-stepped agent-based simulator developed by
Parallel and Distributed Simulation Research Group of the University
of Bologna~\cite{pads}. The tool provides a scalable and customizable
simulation environment where users can define the behavior of the
simulated entities and the data that is exchanged between agents. The
software relies on the GAIA/ART\`IS middleware~\cite{gda-simpat-2017},
which manages communication between logical processes, supporting
parallel and distributed execution, migration of simulated entities,
and load balancing both for computation and communication.

NetLogo is a multi-agent modelling environment based on the Logo
programming language. NetLogo provides an integrated graphical
environment to develop, execute and debug agent-based
simulations. NetLogo models are built around \emph{turtles} (i.e.,
situated agents) that can move around 2D or 3D space, and
\emph{patches} representing portions of space that do not move but can
hold location-specific state data. NetLogo supports real-time plotting
and graphing facilities to display metrics of interest in real
time. NetLogo also supports a form of multilevel modelling through the
LevelSpace extension that allows modelers to construct multilevel
agent-based models within the NetLogo modeling
environment~\cite{hjorth2020levelspace}. LevelSpace allows the
execution and management of multiple models in parallel. However,
execution of models of different types (e.g., a continuous model
within NetLogo) is not supported natively, since the language does not
provide a native way to call external applications. This problem can
be solved by launching NetLogo through
pyNetLogo~\cite{jaxa2018pyNetLogo}, a Python interface to NetLogo,
allowing the developers to start a NetLogo model and to call its
functions from a python script; execution of sub-models is then
delegated to pyNetLogo rather than NetLogo itself.

Finally, the custom-built continuous models are based on the classic
compartmental models, where the population is labelled with a unique
feature, and where usually a set of ordinary differential equations
defines the transition rule between compartments. This approach is
widely used in epidemiology~\cite{brauer2008compartmental}. In our use
cases, we developed the compartmental models in Python, employing the
scipy library~\cite{oliphant2004scipy} to manage the differential
equations.

\begin{figure}[ht]
    \centering\includegraphics[width=\columnwidth]{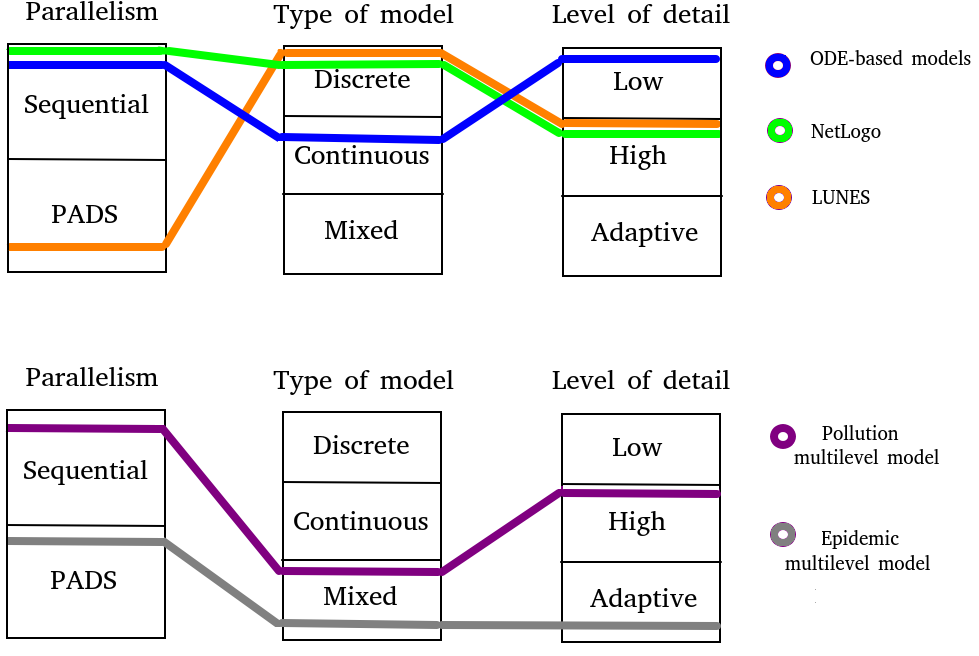}
	\caption{Top: Classification of the simulators used in the case studies. Bottom: classification of the models used in the case studies.}\label{fig:simulators}
\end{figure}

The top of Figure~\ref{fig:simulators} classifies the building blocks
above (NetLogo, LUNES and the scipy continuous model) with respect to
the taxonomy introduced at the beginning of this section. NetLogo is a
sequential, discrete simulator that is suitable for implementing
detailed models, since the behavior of each agent (turtle, in NetLogo
terminology) can be accurately programmed. \ac{LUNES} is a parallel
discrete simulator that, like NetLogo, allows very detailed system
specifications by programming the behavior of agents using the C
language. Finally, our custom continuous model provides coarse systems
specifications using a set of~\aclp{ODE}.

The bottom of Figure~\ref{fig:simulators} classifies the two case
studies that will be described in the next Section. The case studies
implement simple multilevel models to analyze pollution due to
different urban transportation strategies, and the diffusion of an
epidemic. Both are mixed models since continuous and discrete
simulators are used at different levels. The pollution case study is
slightly more detailed and uses a sequential execution policy for all
levels, while the epidemic model makes use of~\ac{PADS} techniques
with the ability to dynamically tune the~\ac{LoD} at run-time.

\section{Case Studies}\label{sec:case-studies}

\begin{figure}[ht]
    \centering
	\includegraphics[width=\columnwidth]{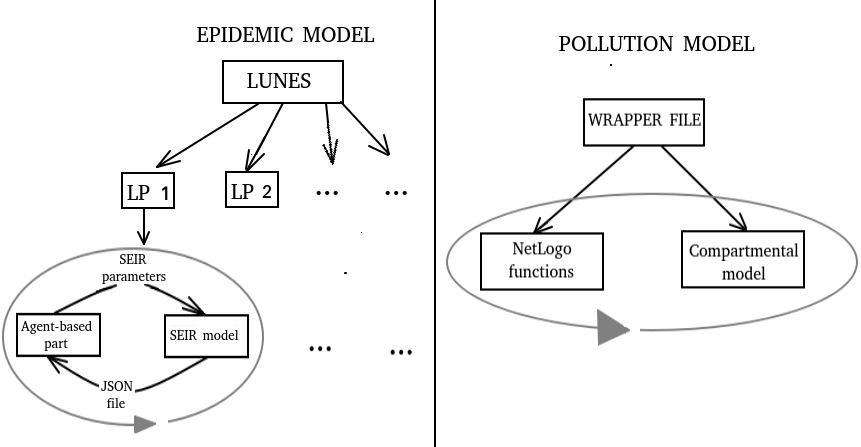}
	\caption{Rationale of the multilevel models for our use cases.}\label{fig:structuring}
\end{figure}

In this section we describe two case studies that demonstrate the
multilevel methodology described above and whose structuring is
schematized in Figure~\ref{fig:structuring}. The case studies are not
intended to be accurate or realistic; they are intended only to
demonstrate that multilevel models can be "natural" description of
scenarios involving human mobility, and how multilevel models can be
implemented in practice. To foster the reproducibility, all the source
code used in this performance evaluation is freely available on
\url{https://github.com/luca-Serena/Multilevel-use-cases} with a Free
Software license.

\subsection{Epidemic Modeling}\label{sec:epidemic}

The first case study is the simulation of an epidemiological
scenario. The model employs two levels, one based on~\ac{LUNES} and in
charge of describing the mobility of individuals, and the other based
on a continuous~\acs{SEIR} model describing the diffusion of the
epidemic through time. A model like this could be used to evaluate the
effectiveness of mobility restrictions to limit the diffusion of an
epidemic. For example, we might compare no-lockdown, full lock-down,
or restrictions based on the job of the individuals where only
essential workers (e.g.,~healthcare personnel, food chain employees,
bus drivers) were allowed to leave their homes.

We consider a non-fatal contagious disease for which permanent
immunity is gained after contagion or immunization. The~\ac{SEIR}
model~\cite{li1995global} is a compartmental model where individuals
are partitioned into four categories:
\begin{itemize}
\item \textit{Susceptible}: individuals that may contract the
  disease. This is the initial state for all the individuals except
  for a single infected individual, the "patient zero".
\item \textit{Exposed}: individuals that contracted the disease but
  are not yet experiencing symptoms.
\item \textit{Infected}: individuals that contracted the disease and
  are experiencing symptoms.
\item \textit{Recovered}: individuals that developed immunity.
\end{itemize}

The dynamics of the~\ac{SEIR} model is governed by the following set
of differential equations:

\begin{equation}
\displaystyle\left\{
\begin{aligned}
\frac{dS}{dt} &= -{\beta IS},\\[6pt]
\frac{dE}{dt} &= \beta IS-\sigma E,\\[6pt]
\frac{dI}{dt} &= {\sigma E}-\gamma I,\\[6pt]
\frac{dR}{dt} &= \gamma I
\end{aligned}
\right.
\end{equation}

\noindent where~$S$, $E$, $I$ and~$R$ are the number of Susceptible,
Exposed, Infected and Recovered individuals, respectively, $\beta$ is
the infection rate, $\gamma$ is the recovery rate and~$\sigma$ is
related to the probability of transmission of the disease.

We entrust~\ac{LUNES} to model the mobility of agents, where each
agent has a position and an epidemiological status; it should be noted
that, for more realistic results, more information such as age, known
health problems, mobility habits or vaccination status can be attached
to agents. We assume that the model consists of several cities; the
population density is higher inside a city, and agents also tend to
move within their own city. Model parameters are the number of cities
and their population, the mobility rates describing how often people
move between cities, and the lockdown policies employed in each
city. For each city, we create a software entity called the "local
coordinator", which is a special simulated entity that has the task of
aggregating the state of all individuals in the city and executing the
continuous model.

The simulation starts with the launch of~\ac{LUNES}, which deals with
the initialization of the virtual environment. In the setup phase, all
the simulated entities are labelled as susceptible, with a single
random individual chosen as the "patient zero". Each agent has a home
location (the city where the associated individual resides), and an
occupation that reflects the frequency of movement and whether the
individual can be classified as an "essential worker".

After the setup phase, the following tasks are performed:
\begin{itemize}
\item The agents send their epidemiological status to the local
  coordinator associated with their location (city).
\item The coordinators aggregate the data and compute the number of
  individuals in each epidemiological state (susceptible, infected,
  ...).
\item The coordinators check if the number of infected individuals is
  above the threshold for triggering lockdown policies, possibly
  restricting the mobility; thereafter, the~\ac{SEIR} model is
  executed using the aggregate values computed above. The various
  continuous models are executed in parallel.
\item The coordinators, based on the results of the~\ac{SEIR} model,
  give instructions to the simulated entities on how to update their
  epidemiological status.
\item Finally, some individuals are moved to a different city with a
  certain probability~$p$.
\end{itemize}

Mobility restrictions have two effects in our model: constraining the
flow of people moving between different areas, and reducing the
contagion rate inside the cities. The model can be made adaptive, for
example changing the parameters of the continuous model such as the
accuracy of the~\ac{ODE} solver by reducing the integration
time-steps. Indeed, we may want to increase the accuracy during the
critical phases of the outbreak when the number of infected
individuals rapidly grows. Also the infection rate could change over
time, e.g., because once the infection is detected, mobility
restrictions could lead to a lower pathogen diffusion rate.

\begin{figure*}[!h]
    \centering
    \begin{subfigure}[t]{0.32\textwidth}
    	\includegraphics[width=1\textwidth]{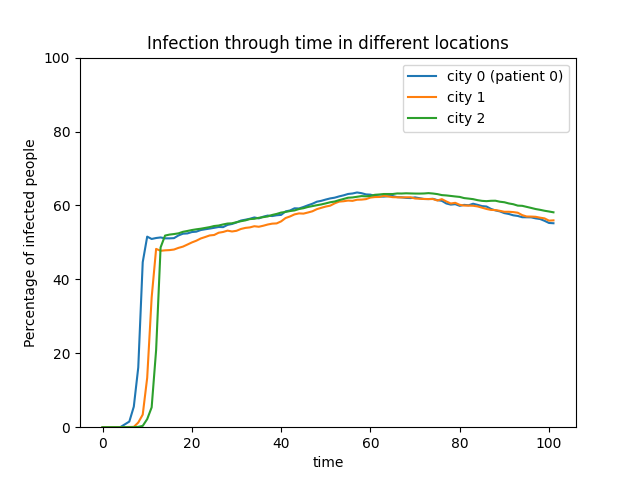}
    	\caption{Number of infected people vs\\ time in different cities.}
    	\label{fig:cities}
    \end{subfigure}
    \begin{subfigure}[t]{0.32\textwidth}
        \includegraphics[width=1\textwidth]{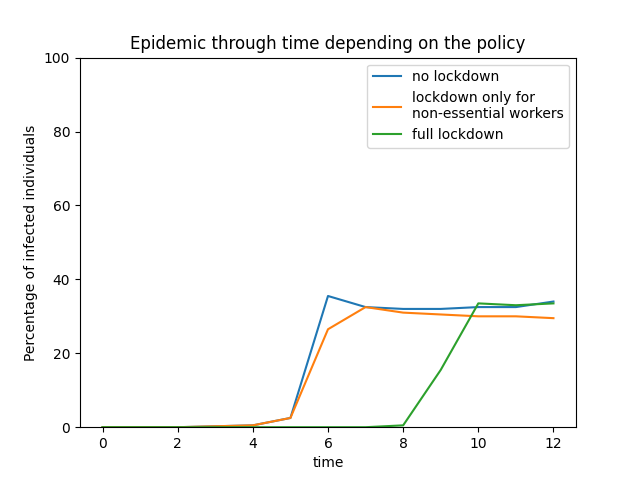}
    	\caption{Diffusion of the epidemic in a\\ location depending on restrictions.}\label{fig:lockdown}
    \end{subfigure}
    \begin{subfigure}[t]{0.32\textwidth}
    	\includegraphics[width=1\textwidth]{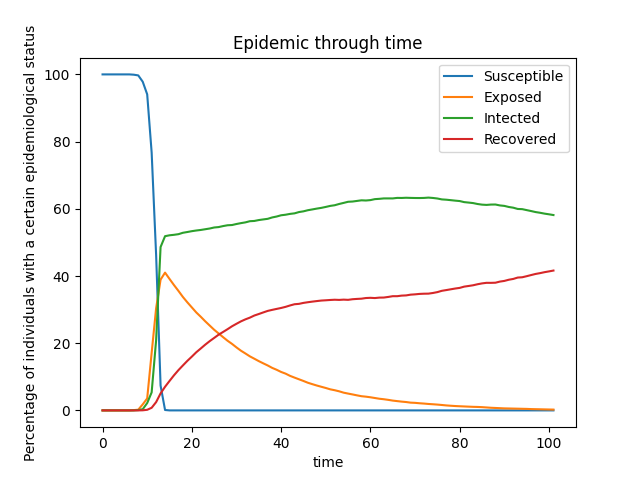}
    	\caption{Number of individuals in each state\\ as a function of time.}\label{fig:SEIR}
    \end{subfigure}
    \caption{}
\end{figure*}

Figure~\ref{fig:cities} shows that the disease propagates rapidly
outside the city of origin. In fact, even assuming that infected
people are not allowed to move outside their home city, exposed
individuals still have no constraints in mobility, and therefore they
will unknowingly diffuse the disease. Figure~\ref{fig:lockdown} shows
that limiting the mobility introduces a delay in the propagation of
the epidemic, but the global impact of the epidemic does not
change. Finally, Figure~\ref{fig:SEIR} displays the progress of the
epidemic inside a single location.

\begin{table}[t]
  \centering%
\begin{tabular}{|c|c|c|}
\hline
   &  \textit{3 Locations} & \textit{6 Locations} \\
\hline
\textit{1000 SEs}  & 50.1 sec & 75.7 sec\\
\textit{10000 SEs} & 49.9 sec & 76.3 sec \\
\hline
\end{tabular}
\caption{Execution time depending on the number of simulated entities and locations. Tests performed on a PC with an 11th gen. Intel Core i5 processor with 16 GBs of RAM running GNU/Linux Ubuntu 20.04.4}\label{tab:perf}
\end{table}

From a performance point of view, as shown in Table~\ref{tab:perf},
the population size has a small impact in terms of execution time,
since it depends mostly on the solution of the differential
equations. On the other hand, the number of locations has a direct
influence on the performance, since despite the greater
parallelization level (under the assumption of using a logical process
for each location) more~\ac{SEIR} models need to be executed, leading
to higher time and memory consumption.

\subsection{Green Mobility to Reduce Pollution}\label{sec:mobility}

The second use case is a model that aims at estimating the amount of
pollutants released into the atmosphere under different vehicular
mobility policies. Vehicles are classified depending on the type of
fuel: gasoline (the most pollutant), electricity (the less pollutant,
although not completely pollution-free since part of the electricity
is still produced using fossil fuels), and LPG (i.e.,~Liquefied
Petroleum Gas) that lies somewhere in between.

We use NetLogo to represent the vehicles. When a vehicle (NetLogo
turtle) moves on a patch (a cell of the discrete simulation space),
the amount of pollution associated with that patch increases by a
quantity that depends on the type of fuel used by the vehicle. Also,
every patch propagates parts of its pollution to the neighboring
patches, to simulate the diffusion of pollutants in the
atmosphere. Periodically, a continuous model is executed to update the
type of vehicles. The continuous model describes the effect of
(dis-)incentives that might be put in place in order to favor some
types of fuel over the others. We defined an ad-hoc compartmental
model based on the following set of differential equations:

\begin{equation}
\left\{
\displaystyle{%
\begin{aligned}
\frac {dP}{dt} &= -\beta P - \sigma P,\\[6pt]
\frac {dL}{dt} &= {\beta P - \gamma L},\\[6pt]
\frac {dE}{dt} &= \sigma P + \gamma L
\end{aligned}}
\right.
\end{equation}\\

\noindent where~$P$, $L$ and~$E$ are the number of petrol, LGP and
electric vehicles, respectively, and $\beta$, $\sigma$ and~$\gamma$
are the transition rates from petrol to LGP, from petrol to electric
and from LGP to electric. For the sake of simplicity, we do not allow
other types of transitions, although they could easily be included by
extending the equations appropriately.

We assume that electric vehicles do not emit pollutants locally but do
contribute slightly to global pollution, since production of energy
may have environmental costs. To model this phenomenon, we
periodically increase the pollution of all patches (even those without
any vehicle) by some small quantity that depends on the number of
electric vehicles in the model.

\begin{figure}[ht]
    \centering%
	\includegraphics[width=.8\columnwidth]{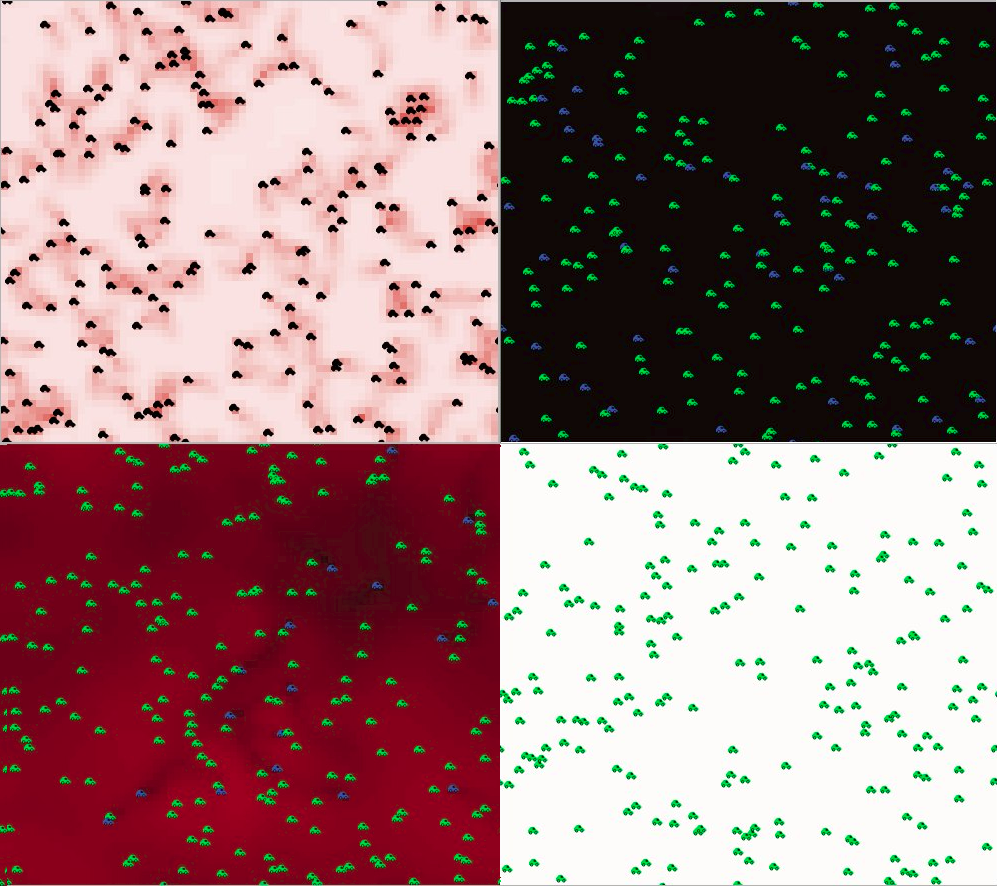}
	\caption{Four snapshots of the NetLogo model at different times. Green turtles = electric vehicles. Blue turtles = LGP-based vehicles. Black turtles = gasoline vehicles. White patches = no pollution. Red patch = medium pollution. Black patches = maximum pollution (best viewed in color).}
	\label{fig:pollution-screen}
\end{figure}

Other important model parameters are (i)~the pollution produced
locally by the vehicles (higher for petrol, lower for LGP; electric
vehicles produce no local pollution but do contribute to global
pollution), (ii)~the "evaporation rate", which is the amount of
pollutants that vanish at each time-step, e.g., because it decays to
inert stuff or is somewhat captured, and (iii)~the initial number of
vehicles of each type. Figure~\ref{fig:pollution-screen} shows a
screenshot of the NetLogo model, where colors represent the type of
vehicles and the amount of pollution in each patch.

To manage the execution of the simulation, we use a wrapper file that
launches NetLogo through the pyNetLogo library and then runs both
NetLogo and the continuous model with the correct parameters.  After
the setup phase, the execution proceeds as follows:

\begin{itemize}
\item each vehicle moves in a random direction, spreading pollutants
  over the visited cells;
\item pollutants diffuse in the atmosphere according to the model
  parameters;
\item every~$n$ steps, where~$n$ is a model parameter, the continuous
  model is called to update the types of vehicles that are moving in
  the simulation.
\end{itemize}

According to the chosen parameters, where we assume that vehicles
running petrol emit as twice pollutants with respect to LPG-based
vehicles and where electricity is produced using zero-emission
technologies, then the model converges towards a zero pollution
situation, meaning that the pollution produced at each time-step is
lower than the evaporation-rate.  In fact, in our scenarios, the
pollution level initially grows, but the curve is gradually reverted
with the switch towards "greener" means of transport. If we assume
that the production of electricity is not exactly zero-emission, the
model converges towards zero pollution more slowly, as long as
$\mathit{NumOfVehicles} \times \mathit{electricityProductionPollution}
< \mathit{EvaporationRate}$.

\begin{figure}[ht]
  \centering%
  \includegraphics[width=.8\columnwidth]{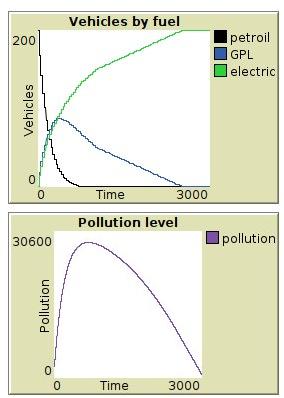}
  \caption{NetLogo real-time plots showing the evolution of the level
    of pollution and the types of vehicles through
    time.}\label{fig:pollution-graphics}
\end{figure}

The time required to converge towards a low pollution state is
dependent on the incentives to purchase less polluting vehicles, as
shown in Figure~\ref{fig:pollution-graphics}. Under the reasonable
assumption that the vast majority of vehicles are initially
petrol-based, pollution grows rapidly at the beginning of the
simulation, but then starts decreasing as more vehicles are replaced
with better ones. Figure~\ref{fig:pollution-screen} shows how the
model progresses over simulated time. Initially (upper left)
petrol-based vehicles start to release pollutants. At simulation
step~$1000$ (upper right) pollution peak is reached, and all patches
are at their maximum contamination level. At simulation step~$2000$
(lower left) most of the vehicles have been replaced with electric
ones, and the environment is getting cleaner. Finally, at simulation
step~$3000$ (lower right) all the patches reached their minimum
pollution level as the green transition is completed.

\section{Conclusions}\label{sec:conclusions}

In this paper we provided some preliminary evidence that multilevel
modelling can be proficiently employed in different simulation
scenarios, in particular those involving human mobility. Despite some
issues regarding the consistency across different models and the need
to coordinate the execution of the various sub-components, there are
indications that the benefits of a multilevel approach outweigh the
cons. First of all, multilevel modeling simplifies the development
phase (i.e.,~different developers implementing separately a different
component of the model) and allows faster execution (i.e.,~certain
sub-models might be launched in parallel). Second, it enables the
integration of existing tools, exploiting already-tested and
task-specific code allowing for faster and more accurate software
development. Finally, describing a complex system at different levels
of detail might bring advantages both computationally (representing a
whole system at a micro level could lead to undue computational
effort) and semantically (having different descriptions of a system).

To support the above points, we provided two simple prototype models
describing the diffusion of a contagious disease, and the diffusion of
pollutants caused by vehicular traffic. In both instances, agent-based
simulators and continuous models have been combined, exchanging
information through JSON files. Our prototyping efforts show that a
multilevel methodology has a great potential for allowing faster
development of complex models that can be easily extended to carry out
more thorough and complex investigations.


\end{document}